\documentclass[aps,pre,twocolumn,groupedaddress,10pt,showpacs,superscriptaddress]{revtex4-1}
\usepackage{amsmath,amssymb}
\usepackage{graphicx,color}

\newcommand{\dif}{\mathrm{d}}%
\newcommand{\fdif}{\operatorname{\delta}}
\newcommand{\Fdif}[2]{\frac{\fdif\!#1}{\fdif\!#2}}
\newcommand{\ii}{i}%
\newcommand{\KR}[1]{\left(#1\right)}%
\newcommand{\Nabla}{\vec{\nabla}}%
\newcommand{\pdif}[2]{\frac{\partial#1}{\partial#2}}%
\newcommand{\ppdif}[2]{\frac{\partial^{2}#1}{\partial#2^{2}}}%
\newcommand{\R}{\mathbb{R}}%

\begin{document}
\title{Microscopic and macroscopic theories for the dynamics of polar liquid crystals}

\author{Raphael Wittkowski}\author{Hartmut L{\"o}wen}
\affiliation{Institut f{\"u}r Theoretische Physik II, Weiche Materie,
Heinrich-Heine-Universit{\"a}t D{\"u}sseldorf, D-40225 D{\"u}sseldorf, Germany}

\author{Helmut R. Brand}
\affiliation{Theoretische Physik III, Universit{\"a}t Bayreuth, D-95540 Bayreuth, Germany}
\date{\today}

\begin{abstract}
We derive and analyze the dynamic equations for polar liquid crystals 
in two spatial dimensions in the framework of classical dynamical density functional
theory (DDFT). Translational density variations, polarization, and quadrupolar order 
are used as order-parameter fields.
The results are critically compared with those obtained using
the macroscopic approach of time-dependent Ginzburg-Landau (GL) equations
for the analogous order-parameter fields.
We demonstrate that for both the microscopic DDFT and the macroscopic GL approach  
the resulting dissipative dynamics can be derived from a dissipation function.
We obtain microscopic expressions for all diagonal contributions and for
many of the cross-coupling terms emerging from a GL approach.
Thus we establish a bridge between molecular correlations and macroscopic modeling
for the dissipative dynamics of polar liquid crystals. 
\end{abstract}


\maketitle


\section{\label{sec:introduction}Introduction}
Understanding the dynamic processes in liquid crystalline systems
from a microscopic point of view is important for designing
smart materials with novel optical and rheological properties.
In fact, the key mechanisms of optical displays and switching
devices are controlled by the dynamic response of liquid
crystals to external stimuli \cite{KimYY2005,AokiOMYT2007,LeeCS2008}.
Therefore, it is highly relevant to understand how these processes change in terms 
of the molecular interactions.
As first example, the switching dynamics of the nematic director upon a change in an
external alignment field \cite{BerardiMZ2008,HaertelBL2010} is one of 
the basic functions in optical displays. Second, in microfluidic devices,
micromixers \cite{GuerraPMC2009} can be tailored by the response of a liquid crystalline
system to a rotating external field.

Nonequilibrium dynamic processes in liquid crystals
are also interesting from a more fundamental point of view since they
involve a nontrivial coupling between translational and orientational
degrees of freedom. Therefore, dynamic macroscopic Ginzburg-Landau (GL) approaches 
have been applied to liquid crystalline systems in order
to obtain the basic dynamic equations on a phenomenological level.
Apart from direct computer simulations of molecular systems \cite{CareC2005,Wilson2007}, 
much less has been done in terms of a microscopic theory. 
Such a microscopic approach which starts from the molecular interactions 
is established by classical density functional theory (DFT)
\cite{Evans1979,Singh1991,Loewen1994a,Loewen2010b,Roth2010}.
DFT can be generalized towards dynamics for colloidal particles in solution 
within the so-called dynamical density functional theory (DDFT)
\cite{MarconiT1999,ArcherE2004,EspanolL2009}, which has been proven
to be a realistic microscopic description for many phenomena
including liquid crystalline dynamics 
\cite{WensinkL2008,HaertelBL2010,HaertelL2010,Loewen2010}.

Static DFT constitutes a framework to derive GL equations
from a microscopic level. The idea is to expand the microscopic one-particle density
in terms of order-parameter fields and to perform a gradient expansion
\cite{LoewenBW1989,LoewenBW1990,OhnesorgeLW1991,Lutsko2006,WittkowskiLB2010,WittkowskiLB2011} 
of the free-energy functional.
This leads to a GL-like theory, which can also be called a phase-field-crystal (PFC) model 
\cite{ElderKHG2002,JaatinenAN2010,MkhontaEG2010,WittkowskiLB2010,WittkowskiLB2011}. 
The phenomenological coupling constants of various terms can thereby be expressed 
in terms of generalized moments of molecular correlation functions.

In this paper, we perform a similar analysis for the \emph{dynamics} of
liquid crystals by using DDFT, on the one hand, and time-dependent GL equations, 
on the other hand.
Therefore, we provide a microscopic basis for time-dependent GL theory and 
derive microscopic expressions for most of the coupling constants. 
We do this in two spatial dimensions for polar liquid crystals
by including translational density variations, polarization, and quadrupolar order
as the basic order-parameter fields.
We further demonstrate that the resulting dissipative dynamics can be
obtained from a dissipation function. 
This DDFT result holds for constant mobilities as well as for  
mobilities depending on the thermodynamic variables. 
Our work opens the way to study various dynamic processes of two-dimensional 
polar liquid crystals recently observed in experiments 
\cite{TabeYNYY2003,FrancescangeliEtAl2009,FrancescangeliS2010}
by further numerical solution of the microscopically justified GL equations.

The paper is organized as follows: in Sec.\ \ref{sec:micd} we derive dynamic equations 
for polar liquid crystals in two spatial dimensions from DDFT.
A macroscopic derivation in terms of time-dependent GL equations and its 
relation to the microscopic dynamics is presented 
in Sec.\ \ref{sec:macd}. We conclude in Sec.\ \ref{sec:conclusions}.

\section{\label{sec:micd}Microscopic derivation of the dynamics}
Our microscopic derivation of the dynamics uses the DDFT equation \cite{WensinkL2008} for collective Brownian motion
of anisotropic uniaxial colloidal particles and is thus based on a static free-energy functional that can be derived from
static DFT theory. 
A perturbative functional, which uses molecular fluid correlations as input, has recently been proposed in our previous work 
\cite{WittkowskiLB2011} for uniaxial polar liquid crystalline particles in two spatial dimensions.
In the following paragraph, we present this functional in a modified form that is more appropriate for our further considerations.

\subsection{Static free-energy functional}
A suitable quantity to describe the state of a system of many interacting anisotropic particles is the 
one-particle density $\rho(\vec{r},\hat{u})$.
It is proportional to the probability density to find a particle with orientation $\hat{u}$ at position $\vec{r}$,
where $\hat{u}=(u_{1},u_{2})=(\cos(\varphi),\sin(\varphi))$ is the orientational unit vector that denotes the orientation of the symmetry axis
of the respective particle and $\vec{r}=(x_{1},x_{2})$ is the center-of-mass position vector. 
In the following, slightly different from Ref.\ \cite{WittkowskiLB2011}, we choose the parametrization
\begin{equation}
\rho(\vec{r},\hat{u}) = 
\bar{\rho}\,\big(1 + \psi(\vec{r}) 
+ P_{i}(\vec{r})\:\!u_{i} + u_{i}\:\!Q_{ij}(\vec{r})\:\!u_{j}\big)
\label{eq:rho}
\end{equation}
with the mean number density $\bar{\rho}$ and the dimensionless order-parameter fields $\psi(\vec{r})$, $P_{i}(\vec{r})$, and $Q_{ij}(\vec{r})$, 
where Einstein's sum convention is used.
These order-parameter fields are the reduced translational density 
\begin{equation}
\psi(\vec{r})=\frac{1}{2\pi\bar{\rho}}\!\int_{\mathcal{S}_{1}}\!\!\!\!\dif\hat{u}\big(\rho(\vec{r},\hat{u})-\bar{\rho}\big)
\end{equation}
with $\mathcal{S}_{1}$ denoting the unit circle. The field $\psi(\vec{r})$ measures translational deviations of $\rho(\vec{r},\hat{u})$ from the mean density $\bar{\rho}$. 
Secondly, the polarization $\vec{P}(\vec{r})$ has the components
\begin{equation}
P_{i}(\vec{r})=\frac{1}{\pi\bar{\rho}}\!\int_{\mathcal{S}_{1}}\!\!\!\!\dif\hat{u}\,\rho(\vec{r},\hat{u})\:\!u_{i}
\end{equation}
and describes the averaged orientation. Finally, the traceless and symmetric nematic tensor with the components
\begin{equation}
Q_{ij}(\vec{r})=\frac{2}{\pi\bar{\rho}}\!\int_{\mathcal{S}_{1}}\!\!\!\!\dif\hat{u}\,\rho(\vec{r},\hat{u})
\Big(u_{i}u_{j}-\frac{1}{2}\delta_{ij}\Big)
\end{equation}
and the Kronecker delta symbol $\delta_{ij}$ describes quadrupolar ordering.
The free-energy functional $\mathcal{F}[\rho(\vec{r},\hat{u})]$ is decomposed as
\begin{equation}
\mathcal{F}[\rho(\vec{r},\hat{u})] = \mathcal{F}_{\textrm{id}}[\rho(\vec{r},\hat{u})] 
+ \mathcal{F}_{\textrm{exc}}[\rho(\vec{r},\hat{u})] 
\end{equation}
into the ideal rotator gas functional
\begin{equation}
\beta\mathcal{F}_{\textrm{id}}[\rho(\vec{r},\hat{u})] =\! 
\int_{\mathcal{A}}\!\!\!\dif\vec{r}\int_{\mathcal{S}_{1}} 
\!\!\!\!\dif\hat{u}\, \rho(\vec{r},\hat{u}) \big(\ln(\Lambda^{2}\rho(\vec{r},\hat{u}))-1\big) 
\label{eq:FidGp}
\end{equation}
with the inverse thermal energy $\beta=1/(k_{\mathrm{B}}T)$, the two-dimensional domain $\mathcal{A}$, and the thermal de Broglie wavelength $\Lambda$
as well as the the excess free-energy functional $\mathcal{F}_{\textrm{exc}}[\rho(\vec{r},\hat{u})]$, which in general is only known approximatively.
Inserting the parametrization \eqref{eq:rho} into Eq.\ \eqref{eq:FidGp} and performing a Taylor expansion up to fourth order in the order-parameter fields
yields to the approximation
\begin{equation}
\beta\mathcal{F}_{\textrm{id}}[\rho(\vec{r},\hat{u})] = F_{\textrm{id}} 
+ \pi\bar{\rho}\int_{\mathcal{A}} \!\!\!\dif\vec{r} \, f_{\mathrm{id}}(\vec{r})
\label{eq:Fida}
\end{equation}
with the local scaled ideal rotator gas free-energy density
\begin{equation}
\begin{aligned}
f_{\mathrm{id}}(\vec{r}) =
&\,\frac{\psi}{4}\big(8-2P^{2}_{i}+2P_{i}Q_{ij}P_{j}-Q^{2}_{ij}\big) \\
&+ \frac{\psi^{2}}{4}\big(4+2P^{2}_{i}+Q^{2}_{ij}\big) - \frac{\psi^{3}}{3} + \frac{\psi^{4}}{6} \\
&+ \frac{P^{2}_{i}}{8}\big(4+Q^{2}_{kl}\big) - \frac{P_{i}Q_{ij}P_{j}}{4} + \frac{P^{2}_{i}P^{2}_{j}}{16} \\
&+ \frac{Q^{2}_{ij}}{4} + \frac{Q^{2}_{ij}Q^{2}_{kl}}{64} \;,
\end{aligned}
\label{eq:Fidb}
\end{equation}
where 
\begin{equation}
F_{\textrm{id}} = 2\pi\bar{\rho}\,A\KR{\ln(\Lambda^{2}\bar{\rho})-1}
\label{eq:FNp}
\end{equation}
is an irrelevant constant and
\begin{equation}
A=\int_{\mathcal{A}}\!\!\!\dif\vec{r}
\end{equation}
is the total area of the domain $\mathcal{A}$.
Considering only second-order terms in the order-parameter fields and performing a gradient expansion up to second order, 
we obtain from the Ramakrishnan-Yussouff approximation \cite{RamakrishnanY1979}
\begin{equation}
\begin{split}
\beta\mathcal{F}_{\mathrm{exc}}[\rho(\vec{r},\hat{u})] = 
-&\frac{1}{2}\int_{\mathcal{A}} \!\!\!\dif\vec{r}_{1}\! \int_{\mathcal{S}_{1}} \!\!\!\!\dif\hat{u}_{1}\!
\int_{\mathcal{A}} \!\!\!\dif\vec{r}_{2}\! \int_{\mathcal{S}_{1}} \!\!\!\!\dif\hat{u}_{2} \,\\
&\!\!\:\!\times c^{(2)}(\vec{r}_{1}-\vec{r}_{2},\hat{u}_{1},\hat{u}_{2}) \\
&\!\!\:\!\times\Delta\rho(\vec{r}_{1},\hat{u}_{1}) \Delta\rho(\vec{r}_{2},\hat{u}_{2})
\end{split}
\label{eq:RY}
\end{equation}
with the direct pair-correlation function 
\begin{equation}
c^{(2)}(\vec{r}_{1},\vec{r}_{2},\hat{u}_{1},\hat{u}_{2})=c^{(2)}(\vec{r}_{1}-\vec{r}_{2},\hat{u}_{1},\hat{u}_{2})
\end{equation}
and the reduced density $\Delta\rho(\vec{r},\hat{u})=\rho(\vec{r},\hat{u})-\bar{\rho}$
for the excess free-energy functional the approximation
\begin{equation}
\beta\mathcal{F}_{\mathrm{exc}}[\rho(\vec{r},\hat{u})] = 
-\frac{1}{2}\!\int_{\mathcal{A}} \!\!\!\dif\vec{r} \, f_{\mathrm{exc}}(\vec{r})
\label{eq:FTE}
\end{equation}
with the local scaled excess free-energy density 
\begin{equation}
\begin{split}
&f_{\mathrm{exc}}(\vec{r}) = 
A_{1}\psi^{2} + A_{2}(\partial_{i}\psi)^{2} + A_{3}(\partial^{2}_{k}\psi)^{2} \\
&\quad\; + B_{1}(\partial_{i}\psi)P_{i} + B_{2}P_{i}(\partial_{j}Q_{ij}) + B_{3}(\partial_{i}\psi)(\partial_{j}Q_{ij}) \\
&\quad\; + C_{1}P^{2}_{i} + C_{2}P_{i}(\partial^{2}_{k}P_{i}) + C_{3}(\partial_{i}P_{i})^{2} \\
&\quad\; + D_{1}Q^{2}_{ij} + D_{2}(\partial_{j}Q_{ij})^{2} \;.
\end{split}
\label{eq:Fexc_a}
\end{equation}
The various coefficients are given by 
{\allowdisplaybreaks\begin{align}%
A_{1} &= 8\,\mathrm{M}^{0}_{0}(1) \;, \\ 
A_{2} &= -2\,\mathrm{M}^{0}_{0}(3) \;, \\
A_{3} &= \frac{1}{8}\,\mathrm{M}^{0}_{0}(5) \;, \\
B_{1} &= 4\big(\mathrm{M}^{1}_{-1}(2)-\mathrm{M}^{0}_{1}(2)\big) \;, \\
B_{2} &= 2\big(\mathrm{M}^{1}_{1}(2)-\mathrm{M}^{2}_{-1}(2)\big) \;, \\
B_{3} &= -\mathrm{M}^{2}_{-2}(3)-\mathrm{M}^{0}_{2}(3) \;, \\
C_{1} &= 4\,\mathrm{M}^{1}_{0}(1) \;, \\
C_{2} &= \mathrm{M}^{1}_{0}(3) - \frac{1}{2}\,\mathrm{M}^{1}_{-2}(3) \;, \\
C_{3} &= -\mathrm{M}^{1}_{-2}(3) \;, \\
D_{1} &= 2\,\mathrm{M}^{2}_{0}(1) \;, \\
D_{2} &= -\mathrm{M}^{2}_{0}(3)
\end{align}}%
as linear combinations of the moments 
\begin{equation}
\mathrm{M}^{m}_{l}(\alpha) = \pi^{3}\bar{\rho}^{2} 
\int^{\infty}_{0}\!\!\!\!\!\!\dif R\, R^{\alpha}\tilde{c}^{(2)}_{l,m}(R) 
\end{equation}
of the Fourier coefficients 
\begin{equation}
\begin{split}
\!\!\!\!\tilde{c}^{(2)}_{l,m}(R)\!=\!\frac{1}{(2\pi)^{2}}\!\!\int^{2\pi}_{0}\!\!\!\!\!\!\!\dif\phi_{\mathrm{R}}\!\!
\int^{2\pi}_{0}\!\!\!\!\!\!\!\dif\phi\, c^{(2)}(R,\phi_{\mathrm{R}},\phi)
e^{-\ii(l\phi_{\mathrm{R}} + m\phi)} \!\!\!\!
\end{split}
\end{equation}
of the direct pair-correlation function $c^{(2)}(R,\phi_{\mathrm{R}},\phi)$, for which the pa\-ra\-me\-tri\-za\-tion  
\begin{equation}
c^{(2)}(\vec{r}_{1}-\vec{r}_{2},\hat{u}_{1},\hat{u}_{2})\equiv
c^{(2)}(R,\phi_{\mathrm{R}},\phi)
\end{equation}
with $\vec{r}_{1}-\vec{r}_{2}=R\hat{u}(\varphi_{\mathrm{R}})$, 
$\hat{u}_{i}=\hat{u}(\varphi_{i})$ for $i=1,2$, 
$\phi_{\mathrm{R}}=\varphi-\varphi_{\mathrm{R}}$, and $\phi=\varphi_{1}-\varphi_{2}$ was used.
Equations \eqref{eq:Fida} and \eqref{eq:FTE} give a local functional of a polar liquid crystalline system
reminiscent of a PFC model \cite{Loewen2010}.

\subsection{\label{subsec:DE}Dynamic equations}
We now derive dynamic equations for $\mathcal{A}=\R^{2}$ for the order-parameter fields 
$\psi(\vec{r},t)$, $P_{i}(\vec{r},t)$, and $Q_{ij}(\vec{r},t)$ from dynamical density functional theory.
DDFT is constructed to describe the Brownian dynamics of colloidal particles in a viscous solvent \cite{Dhont1996,Naegele1996}
via a time-dependent one-particle density field $\rho(\vec{r},\hat{u},t)$. 
This theory was recently extended to anisotropic Brownian particles with orientational degrees of freedom 
\cite{RexWL2007,WensinkL2008,WittkowskiL2011a}. It provides as a starting point for the case of uniaxial particles in 
two spatial dimensions the DDFT equation \cite{WensinkL2008}
\begin{equation}
\begin{split}
\pdif{\rho}{t}(\vec{r},\hat{u},t) \!\:\!\:\!=\!\:\! 
\beta \:\! \Nabla\!\cdot\! 
\Bigg( \!\mathbf{D_{\textrm{T}}}(\hat{u}) \rho(\vec{r},\hat{u},t) 
\Nabla \Fdif{\mathcal{F}[\rho(\vec{r},\hat{u},t)]}{\rho(\vec{r},\hat{u},t)}\! 
\Bigg)& \\ 
+\beta D_{\mathrm{R}}\pdif{}{\varphi} 
\Bigg(\!\rho(\vec{r},\hat{u},t) 
\pdif{}{\varphi} \Fdif{\mathcal{F}[\rho(\vec{r},\hat{u},t)]}{\rho(\vec{r},\hat{u},t)}\! 
\Bigg)&
\end{split}
\label{eq:DDFT}
\end{equation}
with the translational short-time diffusion tensor 
\begin{equation}
\mathbf{D_{\textrm{T}}}(\hat{u})=D_{\parallel}\hat{u}\otimes\hat{u}+D_{\perp}(\mathbf{1}-\hat{u}\otimes\hat{u})\;.
\end{equation}
Here, $D_{\parallel}$ and $D_{\perp}$ are the translational diffusion coefficients for translation parallel and perpendicular 
to the orientation $\hat{u}$, respectively, $D_{\mathrm{R}}$ is the rotational diffusion coefficient, $\otimes$ is the dyadic product, 
and the symbol $\mathbf{1}$ denotes the two-dimensional unit matrix. 
The two terms on the right-hand-side of this DDFT equation for uniaxial particles correspond to pure translation and pure rotation, respectively.
Translational-rotational coupling terms, which are especially relevant for screw-like particles, do not appear in this DDFT equation,
since there is no translational-rotational coupling for uniaxial particles.
Additional terms in the DDFT equation, that regard a possible translational-rotational coupling, would have the same structure as the present terms, 
but with only one gradient and one angular derivative each instead of two gradients or two angular derivatives, respectively \cite{WittkowskiL2011a}.

Following the analysis of Ref.\ \cite{Loewen2010}, the functional derivative $\delta\mathcal{F}/\delta\rho$ in the DDFT equation \eqref{eq:DDFT}
has to be expressed by functional derivatives of the 
free-energy functional with respect to the order-parameter fields $\psi(\vec{r},t)$, $P_{i}(\vec{r},t)$, and $Q_{ij}(\vec{r},t)$,
since we parametrized the one-particle density $\rho(\vec{r},\hat{u},t)$ as well as the free-energy functional 
$\mathcal{F}[\psi,P_{i},Q_{ij}]$ with these order-parameter fields. 
In the following equations, a large number of functional derivatives of the free-energy functional appear. 
Therefore, we shorten the notation by defining the \emph{conjugated order-parameter fields} or \emph{thermodynamic forces}
\begin{equation}
\Xi^{\natural}=\Fdif{\mathcal{F}}{\Xi}\qquad\text{with}\qquad\Xi\in\{\rho,\psi,P_{i},Q_{ij}\}\;.
\end{equation}
Using this notation, the equation
\begin{equation}
\rho^{\natural} =
\frac{1}{2\pi\bar{\rho}}\psi^{\natural} 
+\frac{u_{i}}{\pi\bar{\rho}}P^{\natural}_{i}
+\frac{u_{i}u_{j}}{\pi\bar{\rho}}\:\!Q^{\natural}_{ij} 
\label{eq:LEp}
\end{equation}
follows by functional differentiation.
When performing functional derivatives with respect to $Q_{ij}$ or $Q^{\natural}_{ij}$, 
one has to notice that $Q_{ij}$ as well as $Q^{\natural}_{ij}$ are symmetric and traceless. 
The interdependence of the elements of these tensors leads to more complicated derivatives that respect the 
symmetry properties of these tensors. A very useful equation in this context is
\begin{equation}
\Fdif{Q_{kl}}{Q_{ij}}=\Fdif{Q^{\natural}_{kl}}{Q^{\natural}_{ij}}
=\delta_{ik}\delta_{jl}+\delta_{jk}\delta_{il}-\delta_{ij}\delta_{kl} \;.
\end{equation}
Together with the parametrization \eqref{eq:rho} of the one-particle density, the relation \eqref{eq:LEp} can now be inserted into 
the DDFT equation \eqref{eq:DDFT}.
The dynamic equations for the order-parameter fields are then obtained by an orthogonal projection that separates the 
evolution equations for the particular order-parameter fields from each other.
In doing so, the translational density $\psi(\vec{r},t)$ appears to be conserved, while $P_{i}(\vec{r},t)$ and $Q_{ij}(\vec{r},t)$
are not conserved due to their association with orientational degrees of freedom. 
The dynamic equations can thus be written in the form
{\allowdisplaybreaks
\begin{align}
\begin{split}
\dot{\psi} + \partial_{i} J^{\psi}_{i} &= 0 \;, 
\end{split}\label{dynI}\\
\begin{split}
\dot{P}_{i} + \Phi^{P}_{i}             &= 0 \;, 
\end{split}\label{dynII}\\
\begin{split}
\dot{Q}_{ij} + \Phi^{Q}_{ij}           &= 0 
\end{split}\label{dynIII}%
\end{align}}%
with $\dot{\Xi}=\pdif{\Xi}{t}$ denoting the partial time derivative of the field $\Xi\in\{\psi,P_{i},Q_{ij}\}$
and with the current $J^{\psi}_{i}$ and the quasi-currents $\Phi^{P}_{i}$ and $\Phi^{Q}_{ij}$.
These \emph{dissipative currents} and quasi-currents are given by the expressions
\begin{widetext}
{\allowdisplaybreaks
\begin{align}%
\begin{split}%
J^{\psi}_{i}  = &- \alpha_{1}\big(2(1+\psi)(\partial_{i}\psi^{\natural})+Q_{kl}(\partial_{i}Q^{\natural}_{kl})\big) 
-\alpha_{2}P_{j}(\partial_{i}P^{\natural}_{j}) \\
&-\alpha_{3}\big(2(1+\psi)(\partial_{j}Q^{\natural}_{ij})+P_{i}(\partial_{j}P^{\natural}_{j})+P_{j}(\partial_{j}P^{\natural}_{i})
+Q_{ij}(\partial_{j}\psi^{\natural})\big) \;,
\end{split}\label{eq:Jpsi}\\[4.5pt]%
\begin{split}%
\Phi^{P}_{i}  = &- 2\alpha_{1}\partial_{k}\big(Q_{ij}(\partial_{k}P^{\natural}_{j})+P_{j}(\partial_{k}Q^{\natural}_{ij})\big)
-\alpha_{2}\partial_{k}\big(2(1+\psi)(\partial_{k}P^{\natural}_{i})+P_{i}(\partial_{k}\psi^{\natural})\big) \\[-1pt]
&-\alpha_{3}\Big(2\partial_{i}\big((1+\psi)(\partial_{j}P^{\natural}_{j})\big)+2\partial_{j}\big((1+\psi)(\partial_{i}P^{\natural}_{j})\big)
+\partial_{i}\big(P_{j}(\partial_{j}\psi^{\natural})\big)+\partial_{j}\big(P_{j}(\partial_{i}\psi^{\natural})\big) \\[-3pt]
&\qquad\;\;+2\partial_{j}\big(P_{i}(\partial_{k}Q^{\natural}_{jk})+Q_{jk}(\partial_{k}P^{\natural}_{i})\big)\!\Big)
+\alpha_{4}\big(2(1+\psi)P^{\natural}_{i}+2P_{j}Q^{\natural}_{ij}-Q_{ij}P^{\natural}_{j}\big) \;,
\end{split}\label{eq:PhiP}\\[3pt]%
\begin{split}%
\Phi^{Q}_{ij} = &- 2\alpha_{1}\partial_{k}\big(2(1+\psi)(\partial_{k}Q^{\natural}_{ij})+P_{i}(\partial_{k}P^{\natural}_{j})
+P_{j}(\partial_{k}P^{\natural}_{i})-\delta_{ij}P_{l}(\partial_{k}P^{\natural}_{l})+Q_{ij}(\partial_{k}\psi^{\natural})\big) \\[-1pt]
&-\frac{\alpha_{3}}{2}\Big(4\partial_{i}\big((1+\psi)(\partial_{j}\psi^{\natural})\big)+4\partial_{j}\big((1+\psi)(\partial_{i}\psi^{\natural})\big)
-4\delta_{ij}\partial_{l}\big((1+\psi)(\partial_{l}\psi^{\natural})\big) \\[-2pt]
&\qquad\;\;\,+4\partial_{i}\big(P_{k}(\partial_{j}P^{\natural}_{k})\big)+4\partial_{j}\big(P_{k}(\partial_{i}P^{\natural}_{k})\big)
-4\delta_{ij}\partial_{l}\big(P_{k}(\partial_{l}P^{\natural}_{k})\big) \\[2pt]
&\qquad\;\;\,+\partial_{i}\big(Q_{kl}(\partial_{j}Q^{\natural}_{kl})\big)
+\partial_{j}\big(Q_{kl}(\partial_{i}Q^{\natural}_{kl})\big)
-\delta_{ij}\partial_{l}\big(Q_{km}(\partial_{l}Q^{\natural}_{km})\big) \\
&\qquad\;\;\,+2\partial_{k}\big(Q_{ij}(\partial_{l}Q^{\natural}_{kl})\big)
+2\partial_{k}\big(Q_{kl}(\partial_{l}Q^{\natural}_{ij})\big)\!\Big) \\
&+2\alpha_{4}\big(4(1+\psi)Q^{\natural}_{ij}+P_{i}P^{\natural}_{j}+P_{j}P^{\natural}_{i}-\delta_{ij}P_{l}P^{\natural}_{l}\big)\;,
\end{split}\label{eq:PhiQ}%
\end{align}}%
\end{widetext}
where $\partial_{i}$ are the components of the gradient $\Nabla=(\partial_{1},\partial_{2})$.
Four positive coefficients of whom three are independent appear in these equations. 
With the abbreviation $\lambda=\pi\bar{\rho}/\beta$, they are defined as 
\begin{equation}
\begin{split}
\alpha_{1}&=\frac{D_{\parallel}+D_{\perp}}{8\lambda} \;, \qquad
\alpha_{2}=\frac{D_{\parallel}+3D_{\perp}}{8\lambda} \;, \\
\alpha_{3}&=\frac{D_{\parallel}-D_{\perp}}{8\lambda} \;, \qquad
\alpha_{4}=\frac{D_{\mathrm{R}}}{2\lambda} \;.
\end{split}
\end{equation}
Note that $D_{\parallel}\geqslant D_{\perp}$ holds for all types of uniaxial particles, if the vector $\hat{u}$ for the orientation of the 
symmetry axis is chosen properly
\footnote{The situation described and analyzed here for two spatial dimensions has to be contrasted to the case of three spatial dimensions.
In three spatial dimensions, one has for all rod-like particles, flexible as well as rigid ones, the inequality $D_{\parallel}\geqslant D_{\perp}$, 
while for disk-like particles typically $D_{\parallel}\leqslant D_{\perp}$ applies.}.

At this stage, we emphasize that the DDFT approach \eqref{eq:DDFT} a priori contains only three independent mobility coefficients, 
namely the two translational diffusion coefficients $D_{\parallel}$ and $D_{\perp}$ and the rotational diffusion coefficient $D_{\mathrm{R}}$.
Therefore, all other mobility coefficients for the order-parameter fields can be expressed in terms of these three basic coefficients.
In general, the diffusion coefficients in DDFT are always related to translational or orientational degrees of freedom and not to certain order 
parameters, which appear only with the parametrization of the one-particle density. 
The parametrization of the one-particle density \eqref{eq:rho} in turn does not involve further dissipation coefficients.
This is in sharp contrast to GL theory, where every additional order parameter 
involves at least one new dissipative coefficient -- as will be discussed in Sec. \ref{sec:macd} in more detail.

Since Eqs.\ \eqref{eq:Jpsi}-\eqref{eq:PhiQ} are rather complicated, for numerical calculations a simpler version of these equations 
might be desirable. 
Such a simplification is the \emph{constant-mobility approximation} (CMA), where the one-particle density in the translational and rotational 
mobility terms of the DDFT equation \eqref{eq:DDFT} is approximated by its mean value $\bar{\rho}$:
\begin{equation}
\begin{split}
\pdif{\rho}{t}(\vec{r},\hat{u},t) &=
\beta\bar{\rho}\, \Nabla\!\cdot\! 
\Bigg(\mathbf{D_{\textrm{T}}}(\hat{u}) 
\Nabla \Fdif{\mathcal{F}[\rho(\vec{r},\hat{u},t)]}{\rho(\vec{r},\hat{u},t)}\Bigg) \\ 
&\quad+\beta\bar{\rho}\, D_{\mathrm{R}}\ppdif{}{\varphi} 
\Fdif{\mathcal{F}[\rho(\vec{r},\hat{u},t)]}{\rho(\vec{r},\hat{u},t)}\;.
\end{split}
\label{eq:DDFTpKM}
\end{equation}
With Eq.\ \eqref{eq:DDFTpKM} instead of the DDFT equation \eqref{eq:DDFT}, the following dissipative currents and quasi-currents are obtained:
{\allowdisplaybreaks
\begin{align}%
\begin{split}%
\!\!\!\!\!\!J^{\psi}_{i}  = &- 2\alpha_{1}(\partial_{i}\psi^{\natural})
-2\alpha_{3}(\partial_{j}Q^{\natural}_{ij}) \;, 
\end{split}\label{eq:JpsiCM}\\
\begin{split}%
\!\!\!\!\!\!\Phi^{P}_{i}  = &- 2\alpha_{2}(\partial^{2}_{k}P^{\natural}_{i})
-4\alpha_{3}(\partial_{i}\partial_{j}P^{\natural}_{j})+2\alpha_{4}P^{\natural}_{i} \;,
\end{split}\label{eq:PhiPCM}\\
\begin{split}%
\!\!\!\!\!\!\Phi^{Q}_{ij} = &-4\alpha_{1}(\partial^{2}_{k}Q^{\natural}_{ij})
-2\alpha_{3}\big(2(\partial_{i}\partial_{j}\psi^{\natural})
-\delta_{ij}(\partial^{2}_{k}\psi^{\natural})\big) \!\!\!\!\!\! \\
&+8\alpha_{4}Q^{\natural}_{ij} \;.
\end{split}\label{eq:PhiQCM}%
\end{align}}%
For both the general Eqs.\ \eqref{eq:Jpsi}-\eqref{eq:PhiQ} and the much simpler constant-mobility Eqs.\ \eqref{eq:JpsiCM}-\eqref{eq:PhiQCM},
the explicit forms of the conjugated order-parameter fields $\psi^{\natural}(\vec{r},t)$, $P^{\natural}_{i}(\vec{r},t)$, 
and $Q^{\natural}_{ij}(\vec{r},t)$ result directly from the functional derivatives of Eqs.\ \eqref{eq:Fida} and \eqref{eq:FTE} with respect 
to the order-parameter fields.
These functional derivatives are given by 
{\allowdisplaybreaks
\begin{align}%
\begin{split}%
&\frac{1}{\lambda}
\Fdif{\mathcal{F}_{\mathrm{id}}}{\psi}(\vec{r},t)   =  
2-\frac{P^{2}_{i}}{2}+\frac{P_{i}Q_{ij}P_{j}}{2}-\frac{Q^{2}_{ij}}{4} \\
&\qquad+\frac{\psi}{2}\big(4+2P^{2}_{i}+Q^{2}_{ij}\big) - \psi^{2} + \frac{2}{3}\psi^{3} \;, 
\end{split}\\%
\begin{split}%
&\frac{1}{\lambda}
\Fdif{\mathcal{F}_{\mathrm{id}}}{P_{i}}(\vec{r},t)   =
-\psi\big(P_{i}-Q_{ij}P_{j}\big) + \psi^{2}P_{i} \\
&\qquad+ \frac{P_{i}}{4}\big(4+Q^{2}_{kl}\big) - \frac{Q_{ij}P_{j}}{2} +\frac{P_{i}P^{2}_{j}}{4} \;, 
\end{split}\\%
\begin{split}%
&\frac{1}{\lambda}
\Fdif{\mathcal{F}_{\mathrm{id}}}{Q_{ij}}(\vec{r},t)   =  
\frac{\psi}{2}\big(2P_{i}P_{j}-\delta_{ij}P^{2}_{l}-2\:\!Q_{ij}\big) \\
&\qquad + \psi^{2}Q_{ij} + \frac{P^{2}_{k}}{2}Q_{ij} - \frac{1}{4}\big(2P_{i}P_{j}-\delta_{ij}P^{2}_{l}\big) \\
&\qquad + Q_{ij} + \frac{Q_{ij}Q^{2}_{kl}}{8}
\end{split}%
\end{align}}%
and
{\allowdisplaybreaks
\begin{align}%
\begin{split}%
-2\beta\Fdif{\mathcal{F}_{\mathrm{exc}}}{\psi}(\vec{r},t)  = 
&\,\,2A_{1}\psi-2A_{2}(\partial^{2}_{k}\psi)+2A_{3}(\partial^{2}_{k}\partial^{2}_{l}\psi)\!\!\! \\
&-B_{1}(\partial_{i}P_{i})-B_{3}(\partial_{i}\partial_{j}Q_{ij}) \;,
\end{split}\\%
\begin{split}%
-2\beta\Fdif{\mathcal{F}_{\mathrm{exc}}}{P_{i}}(\vec{r},t)  = 
&\,B_{1}(\partial_{i}\psi) + B_{2}(\partial_{j}Q_{ij}) + 2\:\!C_{1}P_{i} \\
&+2\:\!C_{2}(\partial^{2}_{k}P_{i}) - 2\:\!C_{3}(\partial_{i}\partial_{j}P_{j}) \;,
\end{split}\\%
\begin{split}%
-2\beta\Fdif{\mathcal{F}_{\mathrm{exc}}}{Q_{ij}}(\vec{r},t)  =   
&-B_{2}\big(\partial_{i}P_{j}+\partial_{j}P_{i}-\delta_{ij}(\partial_{l}P_{l})\big) \\
&\hspace{-10mm}-B_{3}\big(2(\partial_{i}\partial_{j}\psi)-\delta_{ij}(\partial^{2}_{l}\psi)\big) + 4D_{1}Q_{ij} \\[2mm]
&\hspace{-10mm}-2D_{2}\:\!\partial_{k}\big(\partial_{i}Q_{kj}+\partial_{j}Q_{ki}-\delta_{ij}(\partial_{l}Q_{kl})\big) \;.\!\!
\end{split}%
\end{align}}%

\subsection{Dissipation function}
In the field of linear irreversible thermodynamics \cite{MartinPP1972,deGrootM1984,Reichl1998}, 
the dissipative parts of the currents and quasi-currents arising in the balance equations for the 
thermodynamic variables (including, for example, hydrodynamic and macroscopic variables) can be derived
from a \emph{dissipation function} $\mathfrak{R}$ which is quadratic in the thermodynamic 
forces. Frequently, one uses equivalently the \emph{entropy production} $\mathfrak{R}/T$
with $T$ denoting the absolute temperature \cite{MartinPP1972,PleinerB1996}. 
The entropy production emerges as a source term in the balance equation
\begin{equation}
\dot{\sigma} + \partial_{i}\:\!j^{\sigma}_{i} = \frac{\mathfrak{R}}{T}
\label{entropybalance}
\end{equation}
for the entropy density $\sigma$, where $\vec{j}^{\sigma}$ is the entropy current density.
The variational derivative of the dissipation function with respect to the thermodynamic
forces gives then the currents and quasi-currents, which are
-- in linear irreversible thermodynamics -- by construction linear
in the thermodynamic forces \cite{MartinPP1972,deGrootM1984,PleinerB1996,Reichl1998}.

Both, dissipation function and entropy production, are maximized close to 
local thermodynamic equilibrium and thus a useful tool in the determination 
of the dissipative currents.
This approach has been applied to a large number of hydrodynamic and macroscopic systems 
\cite{MartinPP1972,deGrootM1984,Forster1989,PleinerB1996,Reichl1998},
but is in general not applicable for active systems and for systems
driven far from equilibrium (compare, for example, Refs.\ \cite{Forster1989,Reichl1998}).
In this more general case, a number of additional conditions must be satisfied
\cite{GrahamH1971a,GrahamH1971b,Risken1972,Graham1974,Risken1996}
in order to obtain a Ljapunov functional. 
Far away from equilibrium, the Ljapunov functional is the analogue of the 
dissipation function of linear irreversible thermodynamics.

More precisely, the dissipative currents and quasi-currents of the dynamic equations 
\eqref{dynI}-\eqref{dynIII} are given by 
{\allowdisplaybreaks
\begin{align}
\begin{split}
J^{\psi}_{i} &= -\Fdif{\mathfrak{R}}{\,(\partial_{i}\psi^{\natural})} \;, 
\end{split}\label{DFI}\\
\begin{split}
\Phi^{P}_{i} &= \Fdif{\mathfrak{R}}{P^{\natural}_{i}} \;, 
\end{split}\label{DFII}\\
\begin{split}
\Phi^{Q}_{ij} &= \Fdif{\mathfrak{R}}{Q^{\natural}_{ij}} \;. 
\end{split}\label{DFIII}%
\end{align}}%
The dissipation function that corresponds to the dissipative currents and quasi-currents \eqref{eq:Jpsi}-\eqref{eq:PhiQ} 
of the general phase-field-crystal (PFC) model is found to be 
\begin{widetext}%
\begin{equation}
\begin{split}
\mathfrak{R}^{(\mathrm{PFC})}=\!\int_{\mathcal{A}} \!\!\!\dif\vec{r} \,\bigg(
&\alpha_{1}\Big((1+\psi)\big((\partial_{i}\psi^{\natural})^{2}+(\partial_{k}Q^{\natural}_{ij})^{2}\big)+
Q_{ij}(\partial_{k}\psi^{\natural})(\partial_{k}Q^{\natural}_{ij})
+Q_{ij}(\partial_{k}P^{\natural}_{i})(\partial_{k}P^{\natural}_{j})\\[-4pt]
&\quad\;\,+2P_{i}(\partial_{k}P^{\natural}_{j})(\partial_{k}Q^{\natural}_{ij})\Big)
+\alpha_{2}\Big(P_{i}(\partial_{j}\psi^{\natural})(\partial_{j}P^{\natural}_{i})+(1+\psi)(\partial_{j}P^{\natural}_{i})^{2}\Big)\\
+\,&\alpha_{3}\Big((\partial_{i}\psi^{\natural})\big(\frac{1}{2}Q_{ij}(\partial_{j}\psi^{\natural})
+P_{i}(\partial_{j}P^{\natural}_{j})+P_{j}(\partial_{j}P^{\natural}_{i})
+2(1+\psi)(\partial_{j}Q^{\natural}_{ij})\big)\\[2pt]
&\quad\;\,+(1+\psi)(\partial_{i}P^{\natural}_{i})^{2}+(\partial_{j}P^{\natural}_{i})\big((1+\psi)(\partial_{i}P^{\natural}_{j})
+Q_{jk}(\partial_{k}P^{\natural}_{i})
+2P_{i}(\partial_{k}Q^{\natural}_{jk})\big)\\[1pt]
&\quad\;\,+\frac{1}{4}Q_{ij}(\partial_{i}Q^{\natural}_{kl})(\partial_{j}Q^{\natural}_{kl})
+\frac{1}{2}Q_{ij}(\partial_{k}Q^{\natural}_{ij})(\partial_{l}Q^{\natural}_{kl})\Big)\\[-1pt]
+\,&\alpha_{4}\Big((1+\psi)\big((P^{\natural}_{i})^{2}+2(Q^{\natural}_{ij})^{2}\big)-\frac{1}{2}Q_{ij}P^{\natural}_{i}P^{\natural}_{j}
+2P_{i}P^{\natural}_{j}Q^{\natural}_{ij}\Big)\!\bigg)\,.
\end{split}
\label{dissgen}
\end{equation}
\end{widetext}%
Together with the dissipative currents and quasi-currents \eqref{eq:Jpsi}-\eqref{eq:PhiQ}, this dissipation function constitutes 
the basic result of this paper.
The dissipation function that corresponds to the currents and quasi-currents \eqref{eq:JpsiCM}-\eqref{eq:PhiQCM} 
of the constant-mobility approximation (CMA) is much simpler and given by 
\begin{equation}
\begin{split}
\mathfrak{R}^{(\mathrm{CMA})}&=\!\int_{\mathcal{A}} \!\!\!\dif\vec{r} \,\Big(
\alpha_{1}\big((\partial_{i}\psi^{\natural})^{2}+(\partial_{k}Q^{\natural}_{ij})^{2}\big)\\
&\qquad+\alpha_{2}(\partial_{k}P^{\natural}_{i})^{2}+\alpha_{4}\big((P^{\natural}_{i})^{2}+2(Q^{\natural}_{ij})^{2}\big)\\[2pt]
&\qquad+2\alpha_{3}\big((\partial_{i}\psi^{\natural})(\partial_{j}Q^{\natural}_{ij})+(\partial_{i}P^{\natural}_{i})^{2}\big)\!\Big)\,.
\end{split}
\label{disscma}
\end{equation}
By construction, both dissipation functions \eqref{dissgen} and \eqref{disscma} are positive.
This is obvious for Eq.\ \eqref{disscma}, but not manifest for Eq.\ \eqref{dissgen}.

\section{\label{sec:macd}Macroscopic approach: Ginzburg-Landau dynamics}
In this section, we investigate the Ginzburg-Landau (GL) dynamics in the vicinity of the phase transitions 
isotropic to polar nematic and isotropic to polar smectic.
In analogy to the previous section, the GL dynamics is discussed for three types of macroscopic variables.
These are the smectic density variation $\rho_{\psi}$, which is closely related  
to the complex scalar $\psi$ often used to describe smectic layering \cite{deGennes1973,ChaikinL1995}, 
the macroscopic polarization $P_{i}$, which becomes important when polar nematic and/or polar smectic phases
are considered \cite{PleinerB1989,BrandPZ2006,BrandCP2009}, and the quadrupolar nematic order parameter $Q_{ij}$, 
that is characteristic of the usual nematic ordering \cite{deGennes1971,deGennesP1995}.

We assume that the local formulation of the first law of thermodynamics, 
the Gibbs-Duhem relation, is valid \cite{deGrootM1984,PleinerB1996,Reichl1998}.
It can be written in the form
\begin{equation}
T \dif\sigma = \dif\varepsilon - \mu \dif\rho  
- \rho_{\psi} \dif\mu_{\psi} - h^{P}_{i} \dif P_{i} - Q_{ij} \dif S_{ij}
\label{gibbs}
\end{equation}
with the absolute temperature $T$, the entropy density $\sigma$, the energy density $\varepsilon$, 
the chemical potential $\mu$, the number density $\rho$, the chemical potential $\mu_{\psi}$ associated with 
the layering $\rho_{\psi}$, the thermodynamic force $h^{P}_{i}$ associated 
with the macroscopic polarization $P_{i}$, and the thermodynamic conjugate $S_{ij}$ of the nematic order parameter $Q_{ij}$.

Throughout the following, we focus entirely on the dissipative dynamics of the variables associated with the additional degrees of
ordering, i.\ e., layering $\rho_{\psi}$, polar order $P_{i}$, and quadrupolar orientational order $Q_{ij}$. 
For the associated dynamic balance equations, we have one dynamic equation each for every hydrodynamic or macroscopic variable. 
These dynamic equations take the form of a conservation law for conserved quantities and are of balance equation type for hydrodynamic 
variables associated with spontaneously broken continuous symmetries and for macroscopic variables such as order parameters close to a 
phase transition.
The dynamic balance equations take thus the form \cite{deGennes1971,MartinPP1972,Forster1989,Khalatnikov1989,PleinerB1996,PleinerLB2002}
{\allowdisplaybreaks
\begin{align}
\begin{split}
\dot \rho_{\psi} + \partial_{i} X^{\psi}_{i} &= 0 \;, 
\end{split}\label{balanceI}\\
\begin{split}
\dot P_{i} + Y^{P}_{i} &= 0 \;, 
\end{split}\label{balanceII}\\
\begin{split}
\dot Q_{ij} + Z^{Q}_{ij} &= 0 \;. 
\end{split}\label{balanceIII}%
\end{align}}%
The currents and quasi-currents $X^{\psi}_{i}$, $Y^{P}_{i}$, and 
$Z^{Q}_{ij}$ are introduced via Eqs.\ \eqref{balanceI}-\eqref{balanceIII}.
Further below, the dissipative part of their structure will be determined 
from the dissipation function $\mathfrak{R}^{(\mathrm{GL})}$. 
There are no reversible currents and quasi-current throughout this paper, 
since flow effects associated with a velocity field $\vec{v}$ or with a density of
linear momentum $\vec{g}$, are generally not considered for the completely overdamped
Brownian dynamics described by DDFT.
We note that the dynamic equation associated with the smectic layering is
of conservation-law type, while the equations for polar and non-polar
orientational order are balance laws.

In the spirit of linear irreversible thermodynamics, we expand
the dissipation function $\mathfrak{R}^{(\mathrm{GL})}$ quadratically in the thermodynamic forces
$\mu_{\psi}$, $h^{P}_{i}$, and $S_{ij}$.
Those in turn have to be determined by taking variational derivatives 
{\allowdisplaybreaks
\begin{align}
\begin{split}
\mu_{\psi} &= \Fdif{\mathcal{F}}{\rho_{\psi}} \;, 
\end{split}\label{forcesI}\\
\begin{split}
h^{P}_{i} &= \Fdif{\mathcal{F}}{P_{i}} \;, 
\end{split}\label{forcesII}\\
\begin{split}
S_{ij} &= \Fdif{\mathcal{F}}{Q_{ij}}
\end{split}\label{forcesIII}%
\end{align}}%
of the suitably chosen generalized potential $\mathcal{F}$ with respect to 
the variables, where $\mathcal{F}$ has been discussed in detail in Refs.\ 
\cite{WittkowskiLB2010,WittkowskiLB2011}.

For the dissipation function associated with the three types of order
considered here, we have to lowest order in the gradients 
\begin{equation}
\begin{split}
\mathfrak{R}^{(\mathrm{GL})}_{0} = \int_{\mathcal{A}} \!\!\!\dif\vec{r} \, &\Big(\frac{1}{2}\gamma_{ijkl}S_{ij}S_{kl} 
+\frac{1}{2}\alpha_{ij}(\partial_{i}\mu_{\psi})(\partial_{j}\mu_{\psi}) \\[-2pt]
&\;+\frac{1}{2}b_{ij}h^{P}_{i}h^{P}_{j} 
+\beta_{ikl}(\partial_{i}\mu_{\psi})S_{kl} \\ 
&\;+\tilde\alpha^{P}_{ij}(\partial_{i}\mu_{\psi})h^{P}_{j}
+\beta^{P}_{ikl}h^{P}_{i}S_{kl}\Big) \;.
\end{split}
\label{eq:RGL}
\end{equation}
It is usual to consider at first only the lowest-order gradient terms in the dissipation function.
Then, one inspects whether contributions containing more gradients are physically relevant. 
For example, one can always add a term containing two more gradients for diagonal terms.
This was done for the diagonal term $\sim\!\gamma_{ijkl}$ in Eq.\ \eqref{eq:RGL}, which leads to a relaxation of the 
nematic order parameter close to the phase transition \cite{deGennes1971}, by the term $\sim\!\tilde{\gamma}_{ijklmn}$ 
in Eq.\ \eqref{Rgrad} further below, which contains two more gradients and is the dissipative analog of the gradient 
energy of the order parameter $Q_{ij}$.

From the dissipation function \eqref{eq:RGL}, we obtain for the dissipative currents and quasi-currents
the expressions
{\allowdisplaybreaks
\begin{align}
\begin{split}
\!\!\!X^{\psi}_{i} &= -\Fdif{\mathfrak{R}^{(\mathrm{GL})}_{0}}{\,(\partial_{i}\mu_{\psi})} 
= -\alpha_{ij} (\partial_{j}\mu_{\psi}) - \tilde\alpha^{P}_{ij} h^{P}_{j} - \beta_{ikl} S_{kl} \,,\!\!\!
\end{split}\label{currentsI}\\
\begin{split}
Y^{P}_{i} &= \Fdif{\mathfrak{R}^{(\mathrm{GL})}_0}{h^{P}_{i}} = 
\tilde\alpha^{P}_{ij} (\partial_{j}\mu_{\psi}) + b_{ij} h^{P}_{j} + \beta^{P}_{ikl} S_{kl} \;,
\end{split}\label{currentsII}\\
\begin{split}
Z^{Q}_{ij} &= \Fdif{\mathfrak{R}^{(\mathrm{GL})}_{0}}{S_{ij}} = 
(\beta_{kij}+\beta_{kji}-\delta_{ij}\beta_{kll})(\partial_{k}\mu_{\psi}) \\[-3pt] 
&\qquad\qquad\;\;\,\:\!+ (\beta^{P}_{kij}+\beta^{P}_{kji}-\delta_{ij}\beta^{P}_{kll})h^{P}_{k} \\[2pt]
&\qquad\qquad\;\;\,\:\!+ (\gamma_{klij}+\gamma_{klji}-\delta_{ij}\gamma_{klmm})S_{kl} \;.
\end{split}\label{currentsIII}%
\end{align}}%
In a truly isotropic phase, one has only two invariants:
the Kronecker delta $\delta_{ij}$ and the totally antisymmetric symbol $\epsilon_{ijk}$.
To preserve the symmetries of such a system,
all the diagonal terms in Eqs.\ \eqref{eq:RGL}-\eqref{currentsIII} contribute, 
while all off-diagonal coupling terms except for one ($\sim\!\alpha^{P}_{ij}$) vanish.
Correspondingly, the property tensors take the form
\begin{equation}
\gamma_{ijkl} = \gamma(\delta_{ik}\delta_{jl} + \delta_{jk}\delta_{il})
\end{equation}
(compare also Ref.\ \cite{Brand1986b}), $\alpha_{ij} = \alpha\delta_{ij}$, and
$b_{ij}=b\delta_{ij}$ for the diagonal terms, and 
$\tilde\alpha^{P}_{ij}=\tilde\alpha^{P}\delta_{ij}$ for the non-vanishing off-diagonal term.

When comparing this result to Eqs.\ \eqref{dissgen} and \eqref{disscma},
we thus arrive at the conclusion that in a time-dependent GL approach 
we have, even to lowest order in the gradients, 
one diagonal dissipative coefficient for each variable entering the dynamics.
This has to be contrasted to the DDFT approach outlined in the last section,
where one has only three independent dissipative transport coefficients
in total (even to higher order in the gradients, compare the discussion
below).
In addition, we find here one off-diagonal contribution, $\sim\!\tilde{\alpha}^{P}$,
which appears to have no analogue in Eqs.\ \eqref{eq:Jpsi}-\eqref{eq:PhiQ}.
By direct comparison, we find explicitly
$\gamma = 2 \alpha_{4}$, $\alpha = 2 \alpha_{1}$, and $b = 2 \alpha_{4}$.

As always, all the dissipative transport coefficients can depend on all
scalar variables in the system including $\rho_{\psi}$ and the temperature $T$.
This general dependence on scalar quantities  
arises partially in the general DDFT result \eqref{dissgen} via the factors
$(1 + \psi)$ instead of $1$, when compared to the CMA. 
If one allows also for a dependence on vector- and tensor-valued variables,
such as the polarization $P_{i}$ and the quadrupolar order $Q_{ij}$,
thus giving up the assumption of strict isotropy, the picture outlined above
changes as follows: the coupling terms between the force associated with $S_{ij}$
and the forces associated with $\partial_{i}\mu_{\psi}$ and $h^{P}_{i}$
can be mediated by the presence of a macroscopic polarization ${P}_{i}$.
In this case, the property tensors $\beta_{ikl}$ and $\beta^{P}_{ikl}$ take the form
\begin{equation}
\begin{split}
\beta_{ikl} &= \beta(\delta_{ik} P_{l} + \delta_{il} P_{k}) \;, \\
\beta^{P}_{ikl} &= \beta^{P}\!(\delta_{ik} P_{l} + \delta_{il} P_{k}) \;.
\end{split}
\label{beta}
\end{equation}
Thus, these dissipative cross-coupling terms can only contribute in the
presence of a macroscopic polarization.
Furthermore, they bring along two additional dissipative coefficients
in a dynamic GL description, while this is not the case for 
the DDFT (compare Eqs.\ \eqref{eq:Jpsi}-\eqref{eq:PhiQ} of Sec.\ \ref{subsec:DE}).
Actually, there appears to be no analogue of the contribution
$\sim\!\beta_{ikl}$ in DDFT, while for the contribution  
$\sim\!\beta^{P}_{ikl}$ we find $\beta^{P} = \alpha_{4}$. 
 
One can also take into account terms containing more gradients
in the dissipation function, as it has been done in the DDFT approach \eqref{dissgen}
and even for the CMA \eqref{disscma}.
To the next order in the gradients, we obtain 
\begin{equation}
\begin{split}
\mathfrak{R}^{(\mathrm{GL})}_{1} = \int_{\mathcal{A}} \!\!\!\dif\vec{r} \, 
&\Big(\frac{1}{2}\tilde \gamma_{ijklmn}(\partial_{m}S_{ij})(\partial_{n}S_{kl}) \\[-3pt]
&\;+\frac{1}{2}\tilde b_{ijkl}(\partial_{k}h^{P}_{i})(\partial_{l}h^{P}_{j}) \\[2pt]
&\;+\tilde \beta_{iklm}(\partial_{i}\mu_{\psi})(\partial_{m}S_{kl}) \\[5pt]
&\;+\tilde\zeta^{P}_{ijk}(\partial_{i}\mu_{\psi})(\partial_{k}h^{P}_{j}) \\[5pt]
&\;+\tilde\xi^{P}_{iklmn}(\partial_{n}h^{P}_{i})(\partial_{m} S_{kl}) \\[3pt]
&\;+\tilde \beta^{P}_{iklm}h^{P}_{i}(\partial_{m}S_{kl})\Big) \;.
\end{split}
\label{Rgrad}
\end{equation}
In a truly isotropic phase, the following picture emerges when Eq.\ \eqref{Rgrad} is analyzed:
the contribution $\sim\!\tilde{\gamma}_{ijklmn}$ is the dissipative analogue
of the gradient energy for the nematic order parameter.
It contains one independent material parameter in two spatial dimensions 
and two parameters in three spatial dimensions. In two spatial dimensions, we have $\tilde{\gamma}\!\sim\!\alpha_{1}$.
The tensor $\tilde{b}_{ijkl}$ has two independent parameters via 
\begin{equation}
\tilde b_{ijkl} = \tilde{b}_{1}\delta_{ik}\delta_{jl}
+ \tilde b_{2} (\delta_{ij}\delta_{kl} + \delta_{jk}\delta_{il}) \;,
\end{equation}
while the tensor $\tilde\beta_{iklm}$ contains one independent parameter
\begin{equation}
\tilde \beta_{iklm} = 
\tilde \beta (\delta_{ik}\delta_{lm} + \delta_{il}\delta_{km}) \;.
\end{equation}
The same applies to the tensor $\tilde\beta^{P}_{iklm}$: 
\begin{equation}
\tilde \beta_{iklm}^{P} = 
\tilde \beta^{P}\!(\delta_{ik}\delta_{lm} + \delta_{il}\delta_{km}) \;.
\end{equation}
All other contributions in Eq.\ \eqref{Rgrad} vanish in a truly isotropic phase. 
Making now an explicit comparison with DDFT, we find 
$\tilde{b}_{1} = 4\alpha_{3}$, $\tilde{b}_{2} = \alpha_{2}$, and $\tilde{\beta} = \alpha_{3}$,
while the contribution $\sim\!\tilde{\beta}^{P}$ has no analogue in DDFT.
The same applies to all contributions which simultaneously contain only one gradient and
are odd in powers of the polarization. 

The contributions $\sim\!\tilde{\zeta}^{P}_{ijk}$ and $\sim\!\tilde{\xi}^{P}_{iklmn}$ 
in Eq.\ \eqref{Rgrad} start to contribute as soon as one allows
a dependence of the property tensors on the polarization $P_{i}$.
These contributions can also be associated with the general DDFT result
given by Eq.\ \eqref{dissgen}.
We note that all the contributions found in Eq.\ \eqref{dissgen}
can also be found in the dynamic GL approach when one allows for a dependence of 
the property tensors on the vector- and tensor-valued variables,
$P_{i}$ and $Q_{ij}$, used here.
However, in contrast to DDFT, these dependencies bring along numerous additional
independent coefficients. 

Thus, we arrive at the conclusion that the DDFT equation \eqref{eq:DDFT} involves three independent
dissipative coefficients in both the general case and the CMA. 
The corresponding terms obtained in the GL framework are associated with nine independent 
coefficients for the analogue of the CMA, i.\ e., for property tensors that do not depend on the variables.
We also note that there are two cross-coupling terms to lowest order in the gradients 
[see Eq.\ \eqref{eq:RGL}], which do not exist in the current DDFT picture.

The overall picture that emerges is therefore the following.
In a dynamic GL approach, there is at least one independent 
dissipative coefficient for every dissipation channel (every order-parameter field)
entering the description. In addition, one finds frequently
dissipative cross-coupling terms that bring along further coefficients 
-- in particular, if one considers a dependence of the dissipative property tensors on
the macroscopic variables.
This can be contrasted to the current DDFT picture, where one has only 
two dissipation channels of diagonal nature, namely translational and rotational diffusion. 
In the present version of DDFT, there are also no independent dissipative cross-coupling terms.
These observations clearly call for a generalization of the current DDFT equation
to incorporate processes that allow for additional dissipation on a microscopic level.

\section{\label{sec:conclusions}Conclusions and possible extensions}
In conclusion, we have proposed both microscopic and macroscopic theoretical descriptions for the dynamics 
of polar liquid crystals in two spatial dimensions. 
The microscopic theory is derived from DDFT, while the macroscopic formulation is based on time-dependent
GL theory. We have done this by including translational density variations, polarization, and quadrupolar order as the 
basic order-parameter fields.
Most but not all phenomenologically possible couplings of GL theory occur also in the DDFT approach. 
These couplings are derived from a microscopic approach and the associated coupling parameters can be
expressed as generalized moments of a molecular correlation function.
We further demonstrated that the whole dynamics can be obtained from a dissipation function. 

Our theoretical framework can be used for a further exploration of various dynamic processes of
polar liquid crystals. This requires numerical solutions of the microscopically justified GL equations following
numerical schemes proposed earlier \cite{vanTeeffelenBVL2009,AchimWL2011}.

For future work, it is challenging to construct a generalized DDFT which explicitly contains the momentum field as appropriate 
for molecular dynamics or systems in flow fields \cite{BraderK2011}. 
This turns out to be much more difficult than the traditional DDFT approach for simple overdamped Brownian dynamics.
But in principle the way of generalization was explored by Tarazona, Marconi and Melchionna \cite{MarconiTCM2008,MarconiM2009,MarconiM2010} 
and by Archer \cite{Archer2006,Archer2009} for molecular dynamics.
An alternative derivation is based on projector techniques \cite{EspanolAZ2009} leading to a hydrodynamical density functional theory 
\cite{RexL2008,EspanolL2009}.
Additional dynamic expressions for a colloidal liquid under shear flow were recently discussed in Ref.\ \cite{BraderK2011}.
Furthermore, a phase-field-crystal model coupled to flow was considered by Voigt and co-workers \cite{AlandLV2010} 
(see also Ref.\ \cite{BorciaBB2008}).
In these extensions, one will presumably obtain non-vanishing 
microscopic expressions for phenomenological dynamic terms caused by the existence of a 
momentum or velocity field.
These come macroscopically mainly in two groups, namely contributions leading to a flow alignment associated with reversible currents 
coupling extensional flow (symmetrized velocity gradients) to orientational degrees of freedom
\cite{Forster1974,BrandP1981a,BrandP1981b,Liu1981,BrandP1982,CladisCB1985,PleinerB1985,PleinerB1996,PleinerLB2002,BrandCP2009}
and coupling terms between extensional flow and variations of the moduli
of nematic, smectic, and columnar order 
\cite{deGennes1971,Liu1979,Brand1986c,deGennesP1995,PleinerB1996,PleinerLB2002}.

\begin{acknowledgments}
We thank Michael Schmiedeberg for helpful discussions.
H.\,L. acknowledges support from the Deutsche For\-schungs\-ge\-mein\-schaft within the science priority program SPP 1296.
H.\,R.\,B. thanks the Deutsche For\-schungs\-ge\-mein\-schaft for partial support of his work through the
Forschergruppe FOR 608 `Nichtlineare Dynamik komplexer Kontinua'.
\end{acknowledgments}

\bibliography{References}
\end{document}